%% file: NSF-Review_arXiv.tex
\documentclass[conference]{IEEEtran}

\usepackage[top=0.75in, bottom=1in, left=0.625in, right=0.625in]{geometry}

\makeatletter
\def\ps@headings{%
\def\@oddhead{\mbox{}\scriptsize\rightmark \hfil \thepage}%
\def\@evenhead{\scriptsize\thepage \hfil \leftmark\mbox{}}%
\def\@oddfoot{}%
\def\@evenfoot{}}
\makeatother
\usepackage{amsmath}
\usepackage{url}
\usepackage{hyperref}
\usepackage{breakurl}
\usepackage{fixltx2e}
\usepackage{graphicx}
\usepackage[usenames]{color}

\ifodd 0
\newcommand{\rev}[1]{{\color{blue}#1}} 
\newcommand{\com}[1]{\textbf{\color{red}(COMMENT: #1)}} 
\newcommand{\comm}[1]{\textbf{\color{green}(#1)}} 
\newcommand{\clar}[1]{\textbf{\color{green}(NEED CLARIFICATION: #1)}}

\else
\newcommand{\rev}[1]{#1}
\newcommand{\com}[1]{}
\newcommand{\comm}[1]{}
\newcommand{\clar}[1]{}
\fi

\begin{document}
\title{Incentives, Quality, and Risks: A Look Into\\
 the NSF Proposal Review Pilot}

\author{\IEEEauthorblockN{Parinaz Naghizadeh}
\IEEEauthorblockA{Department of Electrical Engineering\\
and Computer Science\\
University of Michigan\\
Ann Arbor, Michigan, 48109-2122\\
Email: naghizad@umich.edu}
\and
\IEEEauthorblockN{Mingyan Liu}
\IEEEauthorblockA{Department of Electrical Engineering\\
and Computer Science\\
University of Michigan\\
Ann Arbor, Michigan, 48109-2122\\
Email: mingyan@umich.edu}
}

\maketitle

\begin{abstract} 

The National Science Foundation (NSF) will be experimenting with a new distributed approach to reviewing proposals, whereby a group of principal investigators (PIs) or {\em proposers} in a subfield act as {\em reviewers} for the proposals submitted by the same set of PIs.  To encourage honesty, PIs’ chances for getting funded are tied to the quality of their reviews (with respect to the reviews provided by the entire group), in addition to the quality of their proposals.  Intuitively, this approach can more fairly distribute the review workload, discourage frivolous proposal submission, and encourage high quality reviews. On the other hand, this method has already raised concerns about the integrity of the process and the possibility of strategic manipulation. In this paper, we take a closer look at three specific issues in an attempt to gain a better understanding of the strengths and limitations of the new process beyond first impressions and anecdotal evidence. We start by considering the benefits and drawbacks of {\em bundling} the quality of PIs’ reviews with the scientific merit of their proposals. We then consider the issue of {\em collusion} and favoritism. Finally, we examine whether the new process puts {\em controversial proposals} at a disadvantage. We conclude that some benefits of using review quality as an incentive mechanism may outweigh its drawbacks.  On the other hand, even a coalition of two PIs can cause significant harm to the process, as the built-in incentives are not strong enough to deter collusion. While we also confirm the common suspicion that the process is skewed toward non-controversial proposals, the more unexpected finding is that among equally controversial proposals, those of lower quality get a leg up through this process. Thus the process not only favors non-controversial proposals, but in some sense, mediocrity.  We also discuss possible ways to improve this review process.

\end{abstract}

\input{Intro}

\input{Bonus}

\input{Collusion}
\input{Controversial}

\input{Discussion}

\bibliographystyle{IEEEtran}
\bibliography{IEEEabrv,nsf_sss}

\end{document}

%% file: Intro.tex
\section{Introduction} \label{Intro} 


The National Science Foundation (NSF) makes its funding decisions primarily by using a standard panel review process, whereby a panel of experts are invited to review a set of proposals, discuss their merits at a round table, and make panel/summary assessments and recommendations.  However, the steadily increasing number of submitted proposals to NSF is putting significant stress on this process: A larger number of proposals increases the workload, and a higher workload generally degrades the quality of reviews.  More importantly, the increasing number of proposals in any given field implies a shrinking pool of qualified reviewers in that field as they can no longer contribute to the review process due to conflict of interest considerations; this again leads to concerns that the quality of reviews may degrade. 


This has led the NSF to contemplate alternative peer-review systems, and set out to experiment with a distributed approach to peer-review, detailed in \cite{letter13}. This review process is adapted from the mechanism proposed by Merrifield and Saari in \cite{Merrifield09}, with the main goal to fairly distribute review workload among the proposers themselves\footnote{The terms {\em proposers, applicants, participants} and {\em PIs} will be used interchangeably.}, and increase the quality of the review process. The pilot test is planned for the proposals submitted to the Signal and Sensing Systems (SSS) program by the October 1, 2013 deadline. 
Briefly, this distributed review process proceeds as follows: initially, all submitted proposals are sorted based on their specific subfields, resulting in groups of 25-40 proposals. After collecting conflict of interest information for a group, the main PI for each proposal in a group is required to act as a reviewer for seven of the non-conflicting proposals in his/her group. PIs are given 30 days to complete written reviews, as well as a relative ranking of the proposals in their pile. The partial rankings of the proposals are then compiled into an initial global ranked list. As failure to complete the review task removes a PI's proposal from further consideration, submission of a proposal to this program implies an agreement on the applicants' part to participate in the subsequent review process. In addition, an incentive mechanism is put in place to ensure honest and thorough reviews, whereby the PIs are awarded additional points based on the quality of their reviews judged by how other PIs have rated similar proposals. The initial global ordering is then modified by directly including each PI's bonus points in the assessment of his/her proposal. The program directors make the final decision to award/decline funding based on the modified ranking (see Section \ref{process} for more details).

Despite its benefits, especially in terms of distributing the burden of review, the authors in \cite{Merrifield09}, as well as the academic community, see e.g., \cite{blog1, blog2, blog3}, have expressed concerns about the robustness of this approach. 
Firstly, as the proposed review process is based on a mechanism design approach, without an in-depth study in the corresponding framework, it is not clear whether the inclusion of bonus points provides enough incentive to make truth-telling the (stable) equilibrium of the induced game. 
Without such a study, this process may seem ``unprincipled'' \cite{blog3} and the claims of it being based on mechanism design are deemed ``overblown'' \cite{blog1}. 
Secondly, the best reviewers are not necessarily the best researchers and vice versa \cite{blog2}. Therefore, there is the rather philosophical issue of whether bundling the quality of PIs' reviews with the scientific merit of their proposals will result in discrimination against applicants who are less experienced in the field or less competent reviewers. 
%
Thirdly, while the use of a complex mechanism may help to deter applicants from gaming the outcome, complexity also make it harder for the designers to detect manipulation \cite{blog1}. 

Furthermore, even though the issue of collusion and favoritism is present in many forms of peer-review processes, including the NSF panel-review system, the new approach may present new loopholes and opportunities for exploitation. For one, the reviews are performed anonymously, providing more protection to colluding parties (note that \cite{letter13} not only prevents a PI from finding out who reviewed his/her proposal, as it should, but \rev{with the current specifications,} seems to shield reviewers of the same proposal from each other, which is contrary to current practice). \com{was this point explicit in the NSF letter? if not then we should revise the wording here} \comm{Not explicitly, but there is no specification of whether/how reviewers can contact one another... The original paper says ``...there is no way for presenting killer arguments...'' which can only be because reviewers of the same proposal are not supposed to talk to each other... }
On the other hand, the set of reviewers in each subfield is more or less predictable, in the extreme case leading to a ``hostile takeover of an areas' funding'' \cite{blog1}, where a sufficiently large group of applicants target the same subfield to mutually promote their proposals.

Last but not least, there is the question of whether this procedure will drive all the involved parties (proposers, reviewers, and the funding system) towards favoring non-controversial proposals.  To begin, controversial and high risk proposals are likely to receive mixed reviews. This may ultimately cause them to rank low on average, which then leads the \emph{funding system} to favor less controversial proposals. 
More importantly, as a result of using review quality bonuses, \emph{reviewers} are discouraged from supporting high-risk proposals if they believe this does not reflect the group consensus \cite{Merrifield09, blog1, blog2, blog3}. This problem, which is considered the greatest potential concern by Merrifield and Saari \cite{Merrifield09}, not only drives the reviewers to support less risky proposals, but in turn leads to \emph{proposers} submitting more conservative proposals. 

With these concerns in mind, we set out to examine three specific aspects in the present paper. 
We first analyze the effects of bundling a proposer's performance as a reviewer with the scientific merit of his/her proposal. As such quality bonuses are intended as an incentive mechanism, we start by analyzing their effectiveness in preventing certain forms of dishonest behavior. We compare these positive effects with the two negative effects of discrimination and loss of accuracy induced by their inclusion. Our findings present an overall positive argument in support of such an incentive mechanism, as the downsides appear minor compared to the benefits.   
We then switch focus to the commonly studied issue of collusion, and show that the integrity of a fully decentralized mechanism is quite vulnerable, as the quality bonuses fail to deter even the simplest coalitions.  

Finally, we examine the issue of controversial and high-risk proposals. Note that rewarding PIs with quality bonuses based on their consistency with others results in a \rev{{\em Keynesian beauty contest} \cite{blog1, blog2, camerer97}}, whereby  reviewers strive to conform to the public opinion. Therefore, the first concern regarding controversial proposals, also discussed more frequently in \cite{Merrifield09, blog1, blog2, blog3}, is whether the proposed review process discourages applicants from expressing their honest assessments due to fear of losing quality bonuses. In Section \ref{disc}, we discuss modifications to the mechanism that could prevent such inclinations. We also present a different, perhaps more surprising, aspect of the mechanism when dealing with controversy, only briefly mentioned in \cite{blog1, blog2}. We show that even if reviewers express their honest opinions, the inconsistency among them puts controversial proposals at a disadvantage. Yet the most unexpected and subtle, is the finding that even among equally controversial proposals (but with different quality averaged over mixed reviews), those of lower quality are given an unfair advantage.  Thus the bias toward mediocrity occurs at multiple levels. \footnote{Reviewers' tendency towards favoring non-controversial proposals is what is referred to as \emph{favoring mediocrity} or \emph{conservative mediocrity} in \cite{Merrifield09, blog1, blog2, blog3}. Our findings confirm this effect, but also highlight the tendency to favor the lower quality among the controversial proposals. Therefore, in this paper, ``favoring mediocrity'' will refer to the latter observation.}    

Our study employs a mixture of analysis and simulation. The initial discussion in each section is based on simulation, while a tractable utility function introduced in Section \ref{sec:utility} facilitates analysis for additional insight. {It should also be mentioned that this study looks at the merits of the new review system on its own; it is not meant as a comparison with the current panel system.  For instance, we do not account for the possibility that the quality of participants as reviewers may improve over the current system, partly the rationale behind the new system \cite{letter13}.} 

We end this section with two important notes.  
As part of the research community we very much welcome such an effort by the NSF and the idea of a new and possibly improved review system, 
echoing a sentiment expressed in 
\cite{blog1, blog2, blog3}.  Thus the ultimate goal of this study is to obtain a better understanding of the strengths and limitations of the new approach, so that improvements may be constructed.
{Also worth noting is the fact that though this paper focuses specifically on this new review process, the analysis and conclusions also apply to certain social networking problems, see more discussion in Section \ref{disc}.} 


The rest of the paper is organized as follows. Details of the new review process is presented in Section \ref{process}, followed by a model for the proposers' utility in Section \ref{sec:utility}. We examine the pros and cons of bundling review quality with proposal quality, the process' vulnerability to manipulation by coalitions, and its 
handling of controversial or risky proposals in Sections \ref{Bonus}, \ref{collusion}, and \ref{controversial}, respectively. Section \ref{disc} concludes the paper with suggestions for improvement. 

\section{The NSF Review Pilot: A Distributed Process} \label{process}

The NSF pilot review process \cite{letter13} is based on the distributed review mechanism proposed by Merrifield and Saari in \cite{Merrifield09}, with a few minor modifications. The basic idea is to relegate the task of proposal review to proposers themselves, while taking precautions to deter them from slacking or gaming the process in their own favor. 
A detailed description follows. 

\subsection{Assignment Process}
Upon submission of proposals to the SSS program, the program directors group the proposals based on their specific subfields. Each such subfield, referred to as a \emph{group}, consists of $N$ PIs/proposals, with $N$ typically between 25-40 proposals. 
Each PI is then tasked with reviewing $m$ proposals in his/her group, with which he/she has no personal or organizational conflict of interest (CoI). 
The PIs are given the list of all PIs in their group and are required to declare CoI prior to the review assignment process. 

Each proposal is likewise assigned to $m$ of its permissible reviewers. Therefore, a total of $Nm$ reviews are  collected in each group. The number $m$ is chosen carefully, with the following considerations: (1) it should be small enough to impose a reasonable workload on the reviewers, (2) it should be large enough to discourage the submission of (multiple) low quality proposals, and (3) it should be large enough to extract a reliable overall ranking of the $N$ proposals from the partial rankings (discussed in more detail shortly). 
Accordingly, \cite{letter13} proposes a choice of $m=7$. The assignment of proposals to reviewers is done at random, {subject to different levels of CoI constrains among the reviewers.} 

\subsection{Calculating an Initial Global Ranking} 
\begin{figure}%
\centering
\includegraphics[width=0.8\columnwidth]{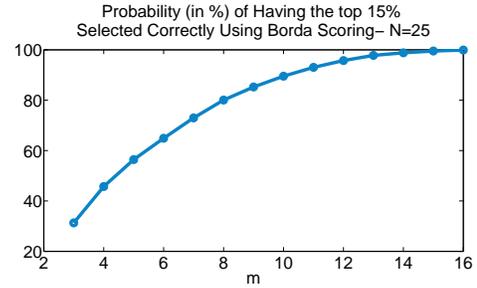}%
\caption{Effect of $m$ on the global list obtained from the partial rankings.}%
\label{eff_m}%
\end{figure}

Each PI is required to provide a written review for each of the $m$ proposals in his/her pile, and to rank the proposals in this set against each other.  The ranking should be based on how the reviewer thinks \emph{the group} will rank the pile, and not on the PIs personal preferences/interests. The goal is to arrive at an objective representation of the communities' view rather than aggregating personal preferences\footnote{This specification in the mechanism, coupled with the quality bonuses detailed in the next subsection, is in fact a main source of concern. Even though these guidelines are intended to encourage honest assessments, they inevitably result in a Keynesian beauty contest, in which reviewers benefit from second guessing one another. See Sections \ref{Intro} and \ref{controversial} for a discussion.}. 
Proposals receive Borda scores \rev{\cite{young95}} of $0$ to $m-1$ in accordance with their position in the PI's ranked list, with the highest ranked proposal receiving $m-1$ points. A proposal's total score is the sum of the $m$ Borda scores assigned to it. Total scores are then normalized into the $[0,1]$ range by dividing by $m(m-1)$. This normalized score is referred to as the Modified Borda Count (MBC) of a proposal. The initial global ordered list is compiled by ranking proposals based on their MBC. 
Figure \ref{eff_m} verifies that even with the choice of a moderate $m$, this initial global ranking is a good representation of the intrinsic merit of the proposals, so that the top proposals are selected with high accuracy. \rev{The results are averages over $10^5$ random proposal assignments.} 

\subsection{Rewarding the Quality of Reviews} 
In order to incentivize PIs to invest time and effort in the review process, and in hopes of deterring manipulation, including defamation and promotion, the proposed review process rewards PIs in accordance with the quality of their reviews. Specifically, a PI's honesty and effort are assessed by comparing his/her proposed ranking with how others have rated similar proposals, using the following absolute deviation from the global ranking as a measure of the quality of PI\textsubscript{$i$}'s reviews: 
\begin{eqnarray}
Q_i &=& \sum_{j=1}^m |{\substack{\text{rank of $j$} \\ \text{in $i$'s list}}} - \substack{\text{rank of $j$ among}\\ \text{these $m$ in global list}}| .
\end{eqnarray}   
The reward to user $i$ is determined by $B_i=2a \frac{Q_{max}-Q_i}{Q_{max}}$. Here, $a:=\frac{MBC_{max}-MBC_{min}}{N}$ is the average difference in MBC between adjacently ranked proposals, and $Q_{max}$ is the worst quality measure possible when $i$'s proposed ranking is exactly the reverse of the global ranking.  
 Consequently, if a PI's submitted ranking is perfectly aligned with the global ranking ($Q=0$), he/she expects a  maximum increase of 2 positions in the final ranked list. 

We would like to point out one of the differences between \cite{letter13}'s adaption of the mechanism proposed in \cite{Merrifield09}. In \cite{letter13}, quality bonuses are used for determining rewards, but not to \emph{weed out} incompetent or malicious reviewers. The authors in \cite{Merrifield09} however consider such possibility. The latter assumption could result in higher accuracy of the final ranking list, especially if there are irrational, incompetent, or arbitrarily malicious participants.   

The final ranking of all $N$ proposals is based on the initial global list, modified with the rewards $B$, so that each PI\textsubscript{$i$}'s final score is determined by $\hat{r}_i = MBC_i + B_i$. Awards are then given to the top ${\cal T}$ proposals (say top 15\% \cite{blog1}) based on the final sorted $\hat{r}$ list. 

\section{A Utility Model and Main Assumptions} \label{sec:utility}
Our study of the distributed review process is based on a mixture of analysis and simulations. We will use the ``probability of getting funded'' as the main metric in all our illustrative numerical experiments. This metric, however, is not amenable to analysis. Therefore, we also introduce the following tractable utility function, which captures important aspects of the real scenario.  

Consider a group consisting of $N$ PIs, each submitting one proposal. Without loss of generality, assume that proposals  are indexed by their intrinsic merit/quality (assumed to exist), with higher indices corresponding to better proposals. Let $\hat{r}_i$ denote the final score of PI\textsubscript{$i$}, $\forall i=1,\ldots, N$. We propose the following model for PI\textsubscript{$i$}'s utility: 
\begin{eqnarray}
u_i = \sum_{j=1, j\neq i}^{N} \frac{\hat{r}_i-\hat{r}_j}{|i-j|^p}~,
\label{eq:utility}
\end{eqnarray}
where $p>0$ is a constant. Firstly, we note that the proposed utility function is increasing in $\hat{r}_i$, reflecting PI\textsubscript{$i$}'s interest in having a higher score than all his/her peers. Secondly, due to the nature of the review process, PI\textsubscript{$i$} is more interested in standing out among his/her closest competitors. To illustrate, consider the following example: if PI\textsubscript{21} can somehow make himself/herself look better in comparison to PI\textsubscript{22}, say by lowering $\hat{r}_{22}$ and increasing $\hat{r}_5$ instead, he/she will be increasing his/her chances of getting funded. This is because the lower quality proposal poses a smaller threat to PI\textsubscript{21}. The term in the denominator of the utility function serves this purpose. 

We would like to note that there are indeed many other possibilities for a utility function. Nevertheless, some  tractable choices such as $u_i = \hat{r}_i$, $u_i = {\hat{r}_i}/(\sum_{j=1}^N \hat{r}_j)$, or $u_i = \sum_{j=1}^{N} (\hat{r}_i-\hat{r}_j)$, despite satisfying the first requirement, fail to capture the latter effect mentioned above. On the other hand, the functions that would capture both effects to varying degrees, e.g. $u_i = \sum_{j=1}^{N} \text{sign}(i-j)|\hat{r}_i-\hat{r}_j|^p$ with $0<p<1$, or (ideally) $u_i = \text{Prob (proposal $i$ is in the top ${\cal T}$)}$, are much harder to analyze, as finding the full probability distribution of $\hat{r}$ is a significant challenge, whereas an analysis based on \eqref{eq:utility} only requires the expected value of the final scores. 

We make the following assumptions throughout the paper. Unless otherwise stated, we assume PIs are equally capable, and have the ability to accurately assess the $m$ proposals in their pile and assign Borda scores according to the proposals' intrinsic merit. A notable exception to this assumption is in Section \ref{Bonus}, when we examine how a PI with a high quality proposal but low review quality may be disadvantaged. Furthermore, we assume reviewers exert their highest level of effort (note that this does not necessarily entail perfect accuracy), and ignore the associated costs. 
Finally, to simplify both the simulation as well as the analysis that follows, we assume there is no CoI except that obviously nobody reviews their own proposal. Therefore, we assume assignments follow a uniform probability distribution. We denote by ${\cal R}_i$ the set of PI/proposal $i$'s reviewers, and by ${\cal A}_i$ the set of proposals assigned to PI\textsubscript{$i$}.

%% file: Bonus.tex
\section{Bundling Review and Proposal Qualities} \label{Bonus} 

The new review proposes incentivizes honest and thorough reviews by incorporating review quality in a proposal's overall assessment. Accordingly, the PIs whose reviews are in sync with other PIs' opinions, are rewarded by having their proposal moved up in the ranked list. 
Despite the possible drawbacks (see Section \ref{Intro}), bundling review and proposal qualities can be beneficial to the review process, as the reward mechanism may be effective in preventing dishonest behavior and unjust reviewing by punishing participants whose opinions are drastically different from others.  More importantly, the mechanism is leveraging the \emph{commodity of interest} \cite{mascolell95} to incentivize PIs' honest cooperation. The main advantage of this approach in contrast to using a \emph{numeraire commodity} (such as taxation), is that it eliminates the need to know the participants' valuation of commodities other than those being allocated (in our case the funding decision). In the current setting, it is reasonable to assume all PIs are interested in getting their proposals funded, whereas their valuation of other commodities such as monetary charges, future privileges or restrictions, and the like, may be difficult or impossible to evaluate.  Below we study the pros and cons of this approach. 

\subsection{Bonuses Can Deter Dishonest Behavior} \label{sec:evil}
We start with an upside of bundling review and proposal qualities, and show
how including quality bonuses prevents certain PIs from an otherwise profitable deviation from honest reporting of reviews. We will focus on one specific form of dishonest behavior, referred to as ``evil'' deviations in \cite{Merrifield09}. An ``evil'' deviator, henceforth simply referred to as a dishonest PI, reverses his/her submitted ranking to bring down the highest ranked proposals, hoping to increase the chances of getting his/her own proposal funded.  Intuitively, this deviation is profitable for PIs of relatively high quality proposals, as slandering other high quality rivals will move them up in the ranking. 

Before proceeding with our discussion, it should be noted that our analysis in this section is not intended to verify that truth-telling is an equilibrium of the game induced by the proposed review mechanism, which is a very interesting subject of future research. There could indeed be other forms of dishonest reporting that would lower competitors' standing while still securing some bonus for the deviator. Our goal in presenting this specific form of dishonest behavior is to highlight the benefits of including quality bonus as part of the final score, at the very least as a means of preventing a plausible type of deviation. 

\subsubsection{Illustrative numerical results} 
\begin{figure}%
\centering
\includegraphics[width=0.8\columnwidth]{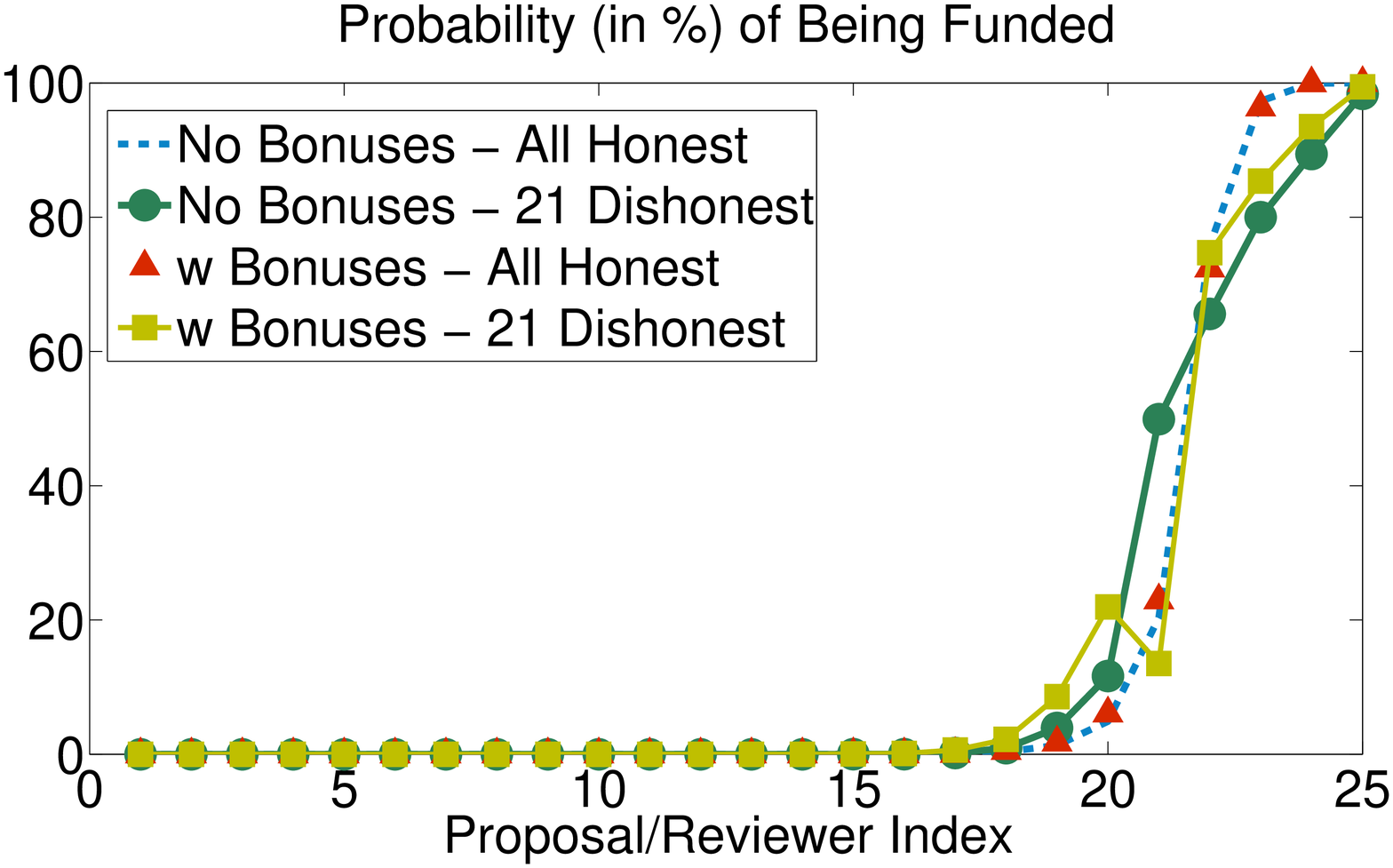}%
\caption{Probability of a proposal getting funded vs intrinsic merit} 
\label{got_in}%
\end{figure}

Set $N=25$, $m=7$, an acceptance rate of $15\%$, and PI\textsubscript{21} as the dishonest PI. Figure \ref{got_in} shows the probability of each proposal getting funded, averaged over $10^5$ random runs with the initial proposal assignment being the random factor. Note that the dishonest PI is chosen to be the one whose own proposal is just on the borderline of being funded.  As also seen in our subsequent results and quite to be expected, the PIs with proposals on the acceptance borderline are more prominently affected by any manipulation of the mechanism or its inherent deficiency. This is because smaller changes in the final scores can change the eventual outcome for these proposals. Therefore, we will often observe a ``reversal of fortune'' at the borderline.  
As shown in Fig. \ref{got_in}, without bonuses, PI\textsubscript{21} can increase his/her chances of being in the top $15\%$ by being dishonest. However, with quality bonuses, a dishonest PI no longer benefits from reverse ranking, since due to loss of bonuses, he/she will fall behind competitors. 

\subsubsection{Analysis} 
An analysis based on the proposed utility model \eqref{eq:utility} is consistent with the above observation. 
If all users $j$ assess and report the Borda scores of their respective sets ${\cal A}_j$ accurately and honestly, the probability that a proposal $i$ receives a Borda score of $k$ from a random reviewer $j$ is given by: 
\[P(j\stackrel{k}{\rightarrow} i) = \frac{{{m-1} \choose k}{{N-m} \choose {i-k-1}}}{{{N-1} \choose {i-1}}}~.\]
Thus, the expected MBC of a proposal $i$ is given by: 
\[E[MBC_i^\text{honest}] = \tfrac{1}{m(m-1)}\sum_{j\in {\cal R}_i} \sum_{k=0}^{m-1} k\cdot P(j\stackrel{k}{\rightarrow} i) = \frac{i-1}{N-1}. \] 
Therefore, if all reports are truthful, PI\textsubscript{$i$}'s utility will be:
\begin{eqnarray}
E[u_i^{T}] &=& \sum_{\substack{j=1\\ j\neq i}}^{N} \frac{E[\hat{r}_i]-E[\hat{r}_j]}{|i-j|^p} \notag\\
 &=& \sum_{j=1, j\neq i}^{N} \frac{\frac{i-1}{N-1}-\frac{j-1}{N-1} + E[B_i-B_j]}{|i-j|^p}\notag\\
 &=& \tfrac{1}{N-1}\sum_{\substack{j=1\\ j\neq i}}^{N} \frac{i-j}{|i-j|^p}~.
\label{eq:truth}
\end{eqnarray} 

We now focus on the effect of reverse ranking. It can be similarly shown that if there are $e$ dishonest PIs in ${\cal R}_i$ reporting a reversed ranking, the expected MBC of the target $i$ is:
\begin{eqnarray*}
E[MBC_i^ \text{ $e$ dishonest}] 
&=& (1-\frac e {m})\frac{i-1}{N-1} + \frac e {m} \frac{N-i}{N-1}~.
\end{eqnarray*}
Assume PI\textsubscript{$i$} is the only dishonest reviewer reporting reversed Borda scores on ${\cal A}_i$. If the quality bonuses are not included in the mechanism, a dishonest PI\textsubscript{$i$}'s utility will be given by:
\begin{eqnarray}
E[u_i^{E}] 
 &=& \sum_{\substack{j=1\\ j\neq i}}^{N} \frac{\frac{i-1}{N-1}-(\frac{j-1}{N-1} +\frac{2}{(N-1)^2}(\frac{N+1}{2}-j))}{|i-j|^p}\notag\\
 &=& \tfrac{1}{N-1}\sum_{\substack{j=1\\ j\neq i}}^{N} \frac{i-j}{|i-j|^p} + \tfrac{2}{(N-1)^2}\sum_{j=1, j\neq i}^{N} \frac{\frac{N+1}{2}-j}{|i-j|^p}\notag\\
 &=& E[u_i^{T}] + \frac{2}{(N-1)^2}\sum_{\substack{j=1\\ j\neq i}}^{N} \frac{\frac{N+1}{2}-j}{|i-j|^p}~.
\label{eq:evil}
\end{eqnarray}
The second term in \eqref{eq:evil} quantifies the gains/losses from dishonest behavior. It is easy to see that this terms is positive for higher $i$ indices, and negative for lower indexed PIs, verifying that reverse reporting is only a profitable deviation for higher quality PIs. Intuitively, degrading other high ranked PIs increases the chances of a good quality proposal for getting funded, as it will look better in comparison with close competitors. Reverse ranking by low quality PIs on the other hand will give more points to their other low ranked rivals, making them look even lower in comparison. In practice, a reviewer cannot necessarily predict where his/her own proposal stands in the global ranking. In this case it may be argued that he/she has little to lose to adopt the reverse ranking, for the proposal gains if it happens to be in the top pile while it really does not matter if it is in the bottom pile as there is not much hope even without the reverse reporting. 

Next, we assume bonuses are awarded based on the quality of reviews. Ideally, when PIs are reporting truthfully they expect a $\frac2N$ bonus added to their initial MBC\footnote{In practice PIs may get less than $\frac2N$ bonus even when they are all truthful, depending on the outcome of the assignment process, see Section \ref{side-effects}.}, while a dishonest PI expects a zero bonus. However, a reversed report by PI\textsubscript{$i$} may result in inaccuracies in the final ranking, in turn making honest PIs lose bonus points.  Assume rather pessimistically that PI\textsubscript{$i$} can alter the final ranking enough to make all honest PIs appear as only moderately accurate reviewers, resulting in a difference in bonus of $\frac1N$\footnote{Based on simulations, honest PIs' loss in {quality bonuses} is much smaller.}.  PI\textsubscript{$i$}'s utility will now be given by:  
\begin{eqnarray}
E[u_i^{E/B}] 
 &=& \sum_{\substack{j=1\\ j\neq i}}^{N} \tfrac{\frac{i-1}{N-1}-(\frac{j-1}{N-1} +\frac{2}{(N-1)^2}(\frac{N+1}{2}-j)) - \frac{1}{N}}{|i-j|^p}\notag\\
 &=& E[u_i^{T}] + \tfrac{1}{(N-1)^2}\sum_{\substack{j=1\\ j\neq i}}^{N} \frac{3-\frac1N-2j}{|i-j|^p}~.
\label{eq:evil_b}
\end{eqnarray}
The second term in \eqref{eq:evil_b} is negative for all $i$. Therefore, once quality bonuses are awarded, reverse ranking will no longer be a profitable deviation. 
 

\subsection{Side Effects of Including Quality Bonuses} \label{side-effects}
We now turn to two possible downsides of this approach.  

\begin{figure}%
\centering
\includegraphics[width=0.75\columnwidth]{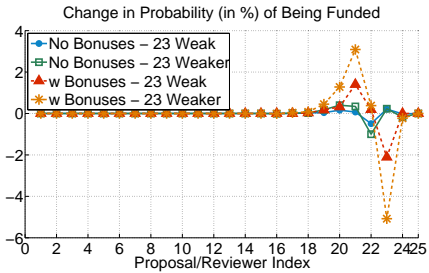}%
\caption{Less competent reviewers have a lower chance of getting funded.}
\label{accuracy0}%
\end{figure}

Firstly, a main premise for justifying the use of quality bonus is that all reviewers are equally competent and accurate when evaluating a proposal. This is however not necessarily true, as the best researchers are not necessarily the best reviewers, and vice versa. To illustrate, set $N=25$, $m=7$, and an acceptance rate of $15\%$. Consider PI\textsubscript{23}, an applicant with a high quality proposal, whose evaluations are affected by a Normally distributed noise of known variance, with a higher variance reflecting a lower level of competentency (recall the review efficiency is determined by how similar PI\textsubscript{23}'s reviews are to the others'). We assume every other PI does a perfect job in ranking the proposals assigned to them. Figure \ref{accuracy0} shows the changes in the probability each proposal gets funded if PI\textsubscript{23} performs less accurately. As shown in Fig. \ref{accuracy0}, when bonuses are added, the chance of PI\textsubscript{23} getting funded decreases as his/her inaccuracy increases. Thus much as expected, those that are less experienced in reviewing will be placed at a disadvantage\footnote{On the up side, quality bonuses can lead to higher quality reviews by incentivizing those who have the expertise to exert high effort, thereby improving the quality of the review process.}.  

\begin{figure}
\centering
\includegraphics[width=0.75\columnwidth]{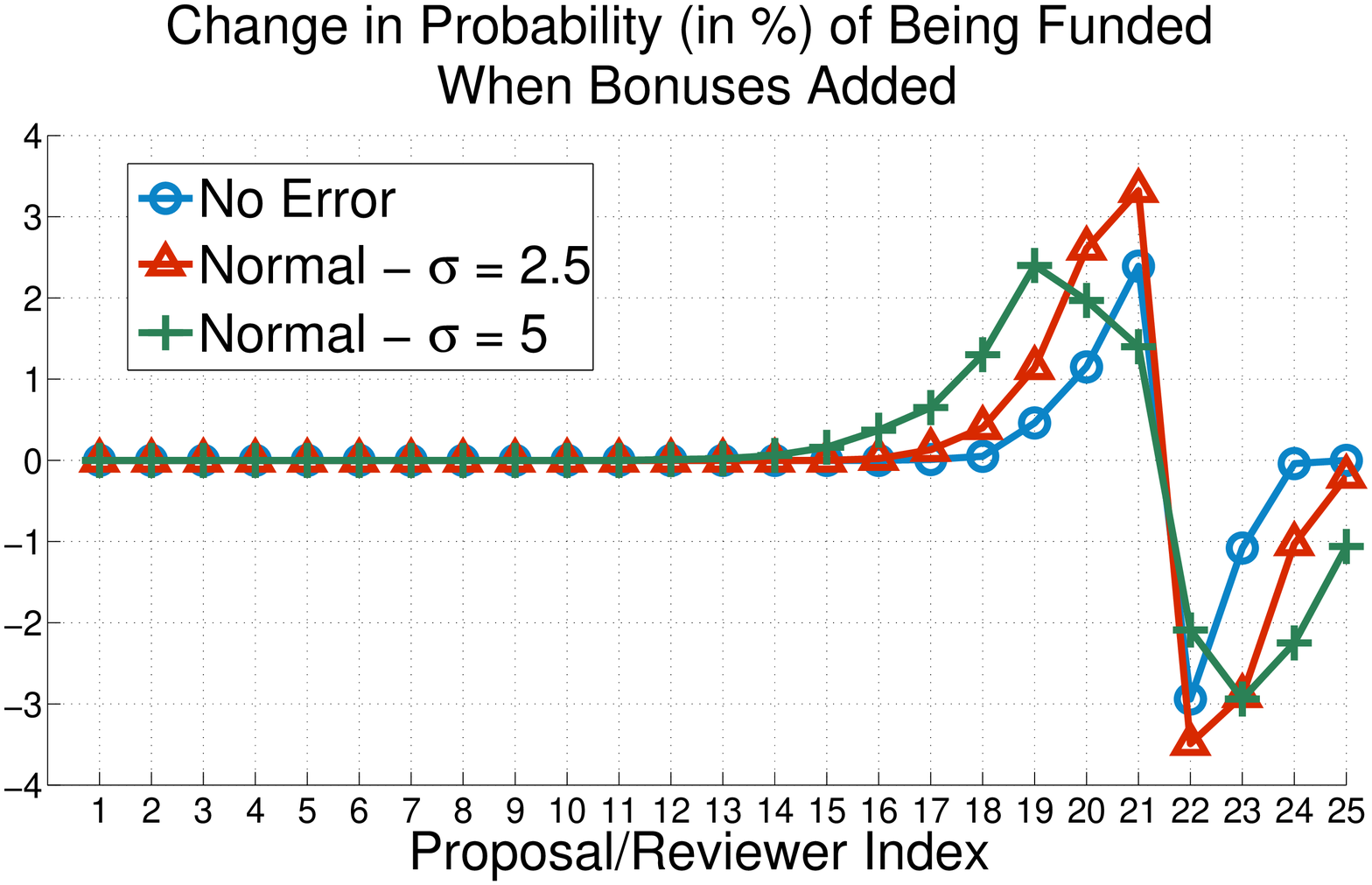}%
\caption{Higher quality proposals are at a disadvantage.}%
\label{accuracy1_diff}%
\end{figure}

Secondly, and perhaps more surprisingly, even if all reviewers are equally capable, accurate, and honest, the addition of quality bonus puts higher quality proposals at a disadvantage. 
To better explain this phenomenon, consider the following counterexample: let $N=25$, $m=7$, and a 15\% funding ratio. Consider two PIs/proposals, $i=22$ who should get funded, and $j=21$ who is right on the borderline. Now assume the outcome of the proposal assignment process is rather {\emph{skewed}}, in the following sense: an average quality proposal, say $15$, happens to fall into multiple batchs of low quality proposals, e.g. ${\cal R}_k = \{3,8,\ldots,15\}$. Due to the structure of Borda scoring, reviewers are required to assign a score of $6$ to the best proposal in their pile regardless of its intrinsic merit. Therefore, the skewed assignment could very well lead to a high global rank for proposal $15$. Now assume $i$ is assigned a more uniform set of proposals, say ${\cal R}_i = \{3,\ldots, 15, \ldots, 23\}$, and assigns the Borda scores $0,\ldots, 6$ to these proposals. It is easy to see that $i$ will get a low quality bonus based on such ranking, even though he/she has been honest and accurate. If another PI, say PI\textsubscript{$j$} has a better quality bonus, he/she may end up getting funded instead of PI\textsubscript{$i$}. 
This observation is reflected in Fig. \ref{accuracy1_diff}, which shows the change in the probability each proposal is funded after the inclusion of quality bonuses, averaged over $10^5$ random proposal assignments.  Figure \ref{accuracy1_diff} refutes the claim in \cite{letter13} that: ``$\ldots$ if all reviewers do an excellent job of ranking the proposals they review, all PIs' proposals will be moved up equally, which means that the ranking will not be changed $\ldots$''.   
Also, following a similar argument, this situation is exacerbated when all PIs' reviews are affected by a Normally distributed noise, as illustrated in Fig. \ref{accuracy1_diff}.  


\begin{figure}%
\centering
\includegraphics[width=0.75\columnwidth]{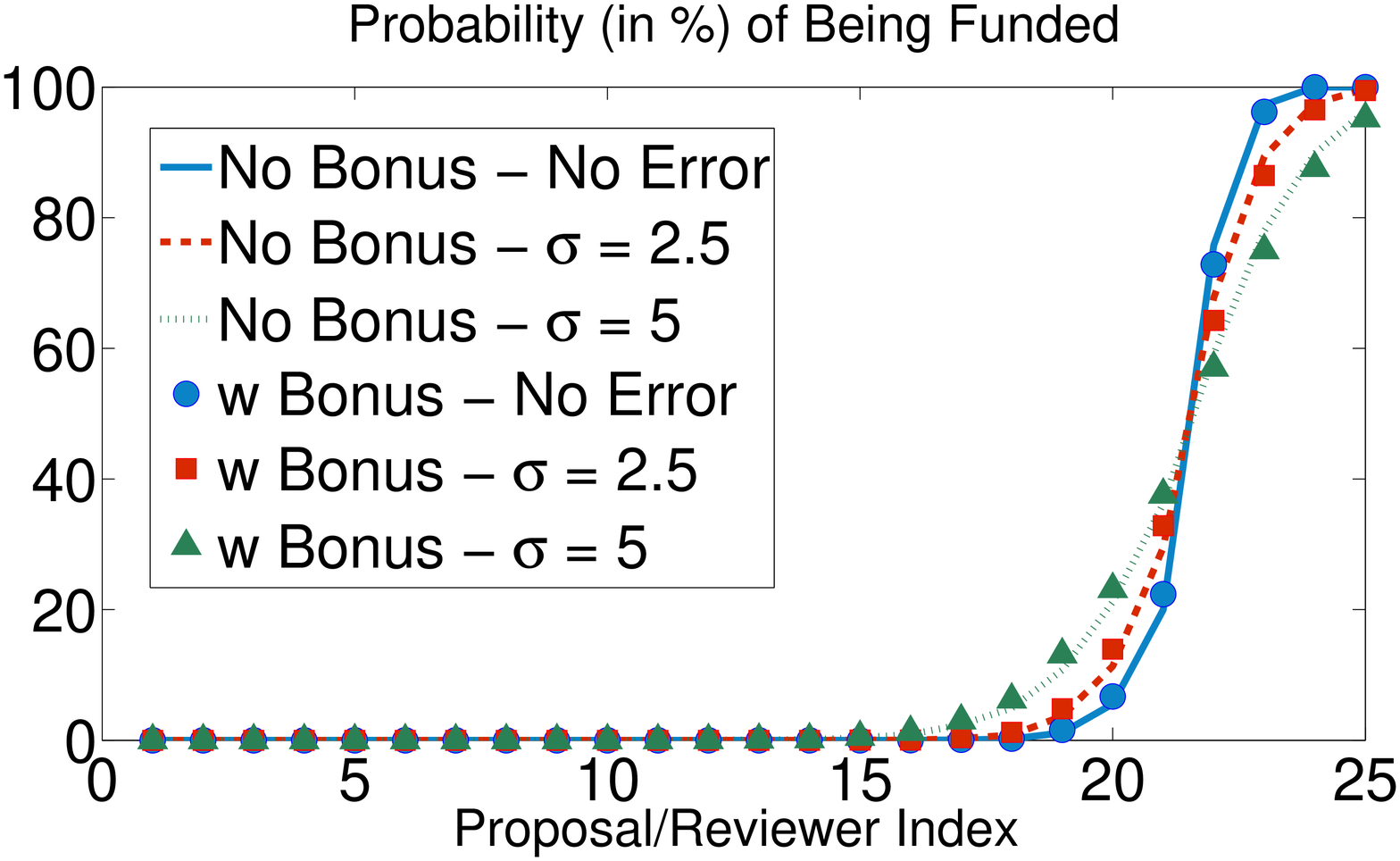}%
\caption{Higher quality proposals still have a higher chance of getting funded.}%
\label{accuracy1}%
\end{figure}

It should be noted that PI\textsubscript{22} has a higher probability of being funded than PI\textsubscript{21}, with or without the inclusion of quality bonus, as shown in Fig. \ref{accuracy1}. {That said, the addition of bonus does  make higher quality proposals' outcome more sensitive to perturbations to the system, e.g., the funding ratio.}  

\begin{figure}
\centering
\includegraphics[width=0.75\columnwidth]{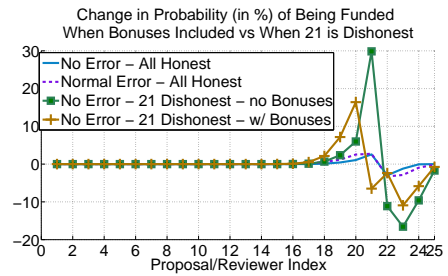}%
\caption{Dishonest ranking does more damage than quality bonuses.}%
\label{accuracy1evil}%
\end{figure} 

\subsection{A Comparison}
{We now compare the quality bonuses' merit in preventing certain malicious intent (e.g., submitting reversed ranking) and its side effects shown above.}   
This is illustrated in Fig. \ref{accuracy1evil}, where $N=25$, $m=7$, acceptance rate is $15\%$, {PI\textsubscript{21}} is dishonest, and all PIs are equally accurate. 
\rev{Numerically, based on Fig. \ref{accuracy1evil}, the dishonest PI can alter his/her chances of getting funded by as much as 30\% when quality bonuses are not used (the largest jump in Fig. \ref{accuracy1evil}), whereas once quality bonuses are included, PI\textsubscript{21} expects a $\sim8\%$ lower chance of being funded if dishonest, and is thus deterred from dishonest reporting. On the other hand, the inclusion of quality bonuses would only result in an inaccuracy of $\sim4\%$ in the probabilities of getting funded ({the smaller bump at index 21, see also Fig. \ref{accuracy1_diff}}). } 
\com{Will this re-wording make it clear?}

Comparing these positive and negative effects, we conclude that direct inclusion of quality bonuses in the assessment of proposals may very well benefit the review process. 
Nevertheless, it is worthwhile to look into incentive methods that use either review quality bonuses or alternative metrics as a numeraire commodity (rather than the commodity of interest). See Section \ref{disc} for a discussion. 


%% file: Collusion.tex
\section{Collusion} \label{collusion} 

We now move on to the issue of collusion and personal favors. 
Unlike the issue of quality bonuses which is specific to the proposed process, collusion can arise in many review systems, including the traditional panel based reviews. 
However, certain features of the new review process may make collusion easier to conduct and harder to detect.  The first is the fact that under the new process a reviewer/PI  gets to see the entire list of other reviewers/PIs in the same group (in order to declare CoI) so it may be easier for someone to determine whether there are opportunities of collusion, whereas in the current panel system reviewers may not find out the identity of the other reviewers until they arrive at the panel.  Secondly, since the reviewers are also PIs, there is the opportunity for direct and instantaneous quid pro quo between two (or more) PIs if the assignment happens to match them up such that they review each other's proposals; by contrast, in the current system the reviewers and PIs are two disjoint sets, thus a favor done (by a reviewer to a PI) cannot be repaid until much later, if ever, and very likely through a different channel as the PI receiving the favor cannot guarantee to later serve on a panel that includes the reviewer's proposal.  Last but not least, under the new process the reviewers appear to enjoy a higher level of anonymity which may provide more protection to colluding parties and makes it harder to detect\footnote{It must to be mentioned that such anonymity also protects honest but different opinions, whereas the current panel system may allow some opinion to trump others.}. 
  
In this section, we focus on the case where there are only two PIs colluding, PI\textsubscript{i} and PI\textsubscript{j}. We consider two possible arrangements between colluding PIs. With \emph{One-Sided} favors, only PI\textsubscript{i} is favoring the other, in case they are matched, as follows: he/she assigns the highest Borda score $m-1$ to PI\textsubscript{j}, and then assigns reverse Borda scores of $m-2$ to $0$ to the remaining proposals, with $0$ points for the highest ranked proposal. This means PI\textsubscript{j} is always benefiting from the collusion (with an increased expected score and slandered rivals), while PI\textsubscript{i} may be putting itself at risk (
{for receiving a low quality bonus}). 
Therefore, one-sided favors reflect a PI's gains with the help of a single ally without the need for immediate payback. 
The second type of arrangement is \emph{Reciprocal} favors: PIs both plan to favor one another in case they are matched, but with a more conservative collusion strategy: if matched they assign $m-1$ to their ally, and then truthfully assign Borda scores of $0$ to $m-2$ to the remaining proposals. Thus, the two PIs are expecting both gains and losses from such an arrangement.  

\subsection{Illustrative Numerical Results} 
For the following simulation, we fix $N=25$, $m=7$ and an acceptance rate of $15\%$. Furthermore, we focus on the colluding parties who have a low chance of getting funded, i.e., those with intrinsic merit of 21 or lower. To see whether collusion is profitable, we look at the changes in probability of a proposal getting funded, with or without collusion. 
We further consider the two cases where a relatively high quality proposal teams up with either a high or a low quality proposal. 

In what follows, the first two cases correspond to one-sided favors, and consider the changes in probability of getting funded assuming PI\textsubscript{i} is one of PI\textsubscript{j}'s reviewers. This event itself occurs with probability ${m}/{N_j}$, where $N_j< N$ is the number of PIs with whom $j$ does not have CoI. When $N$ or $N_j$ are small, PI\textsubscript{j} is more likely to be matched with an ally. For reciprocal favors, we consider the \emph{a priori} probabilities of getting funded, i.e., the PIs do not know whether they will end up being the other's reviewer.  

\begin{figure}%
\begin{minipage}[t]{0.45\linewidth}
\includegraphics[width=\columnwidth]{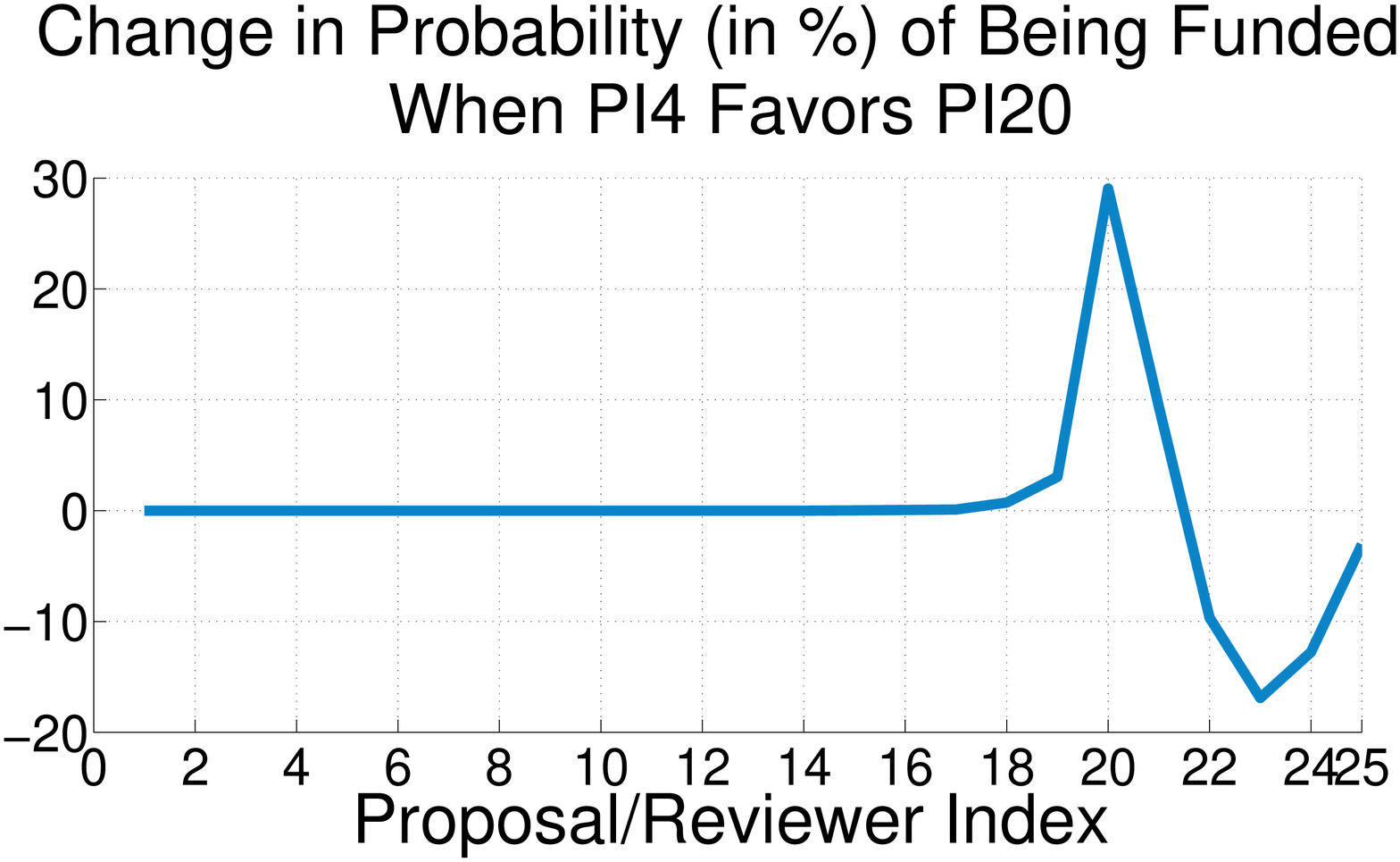}%
\caption{One Sided Favors - low quality favoring high quality}%
\label{1S-HL}%
\end{minipage}
\hspace{0.5cm}
\begin{minipage}[t]{0.45\linewidth}
\includegraphics[width=\columnwidth]{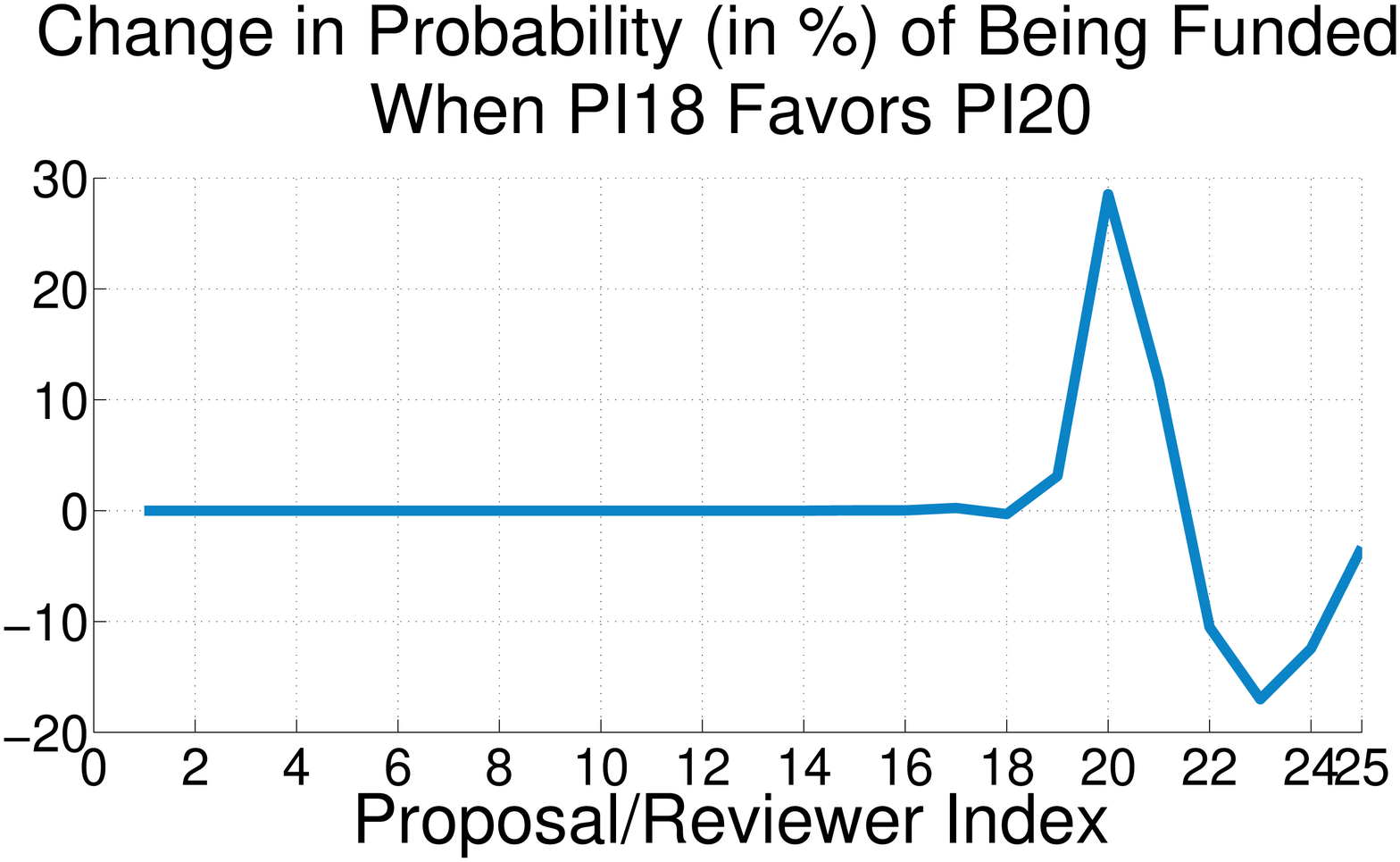}%
\caption{One Sided Favors - high quality favoring high quality}%
\label{1S-2H}%
\end{minipage}

\vspace{0.5cm}

\begin{minipage}[t]{0.45\linewidth}
\includegraphics[width=\columnwidth]{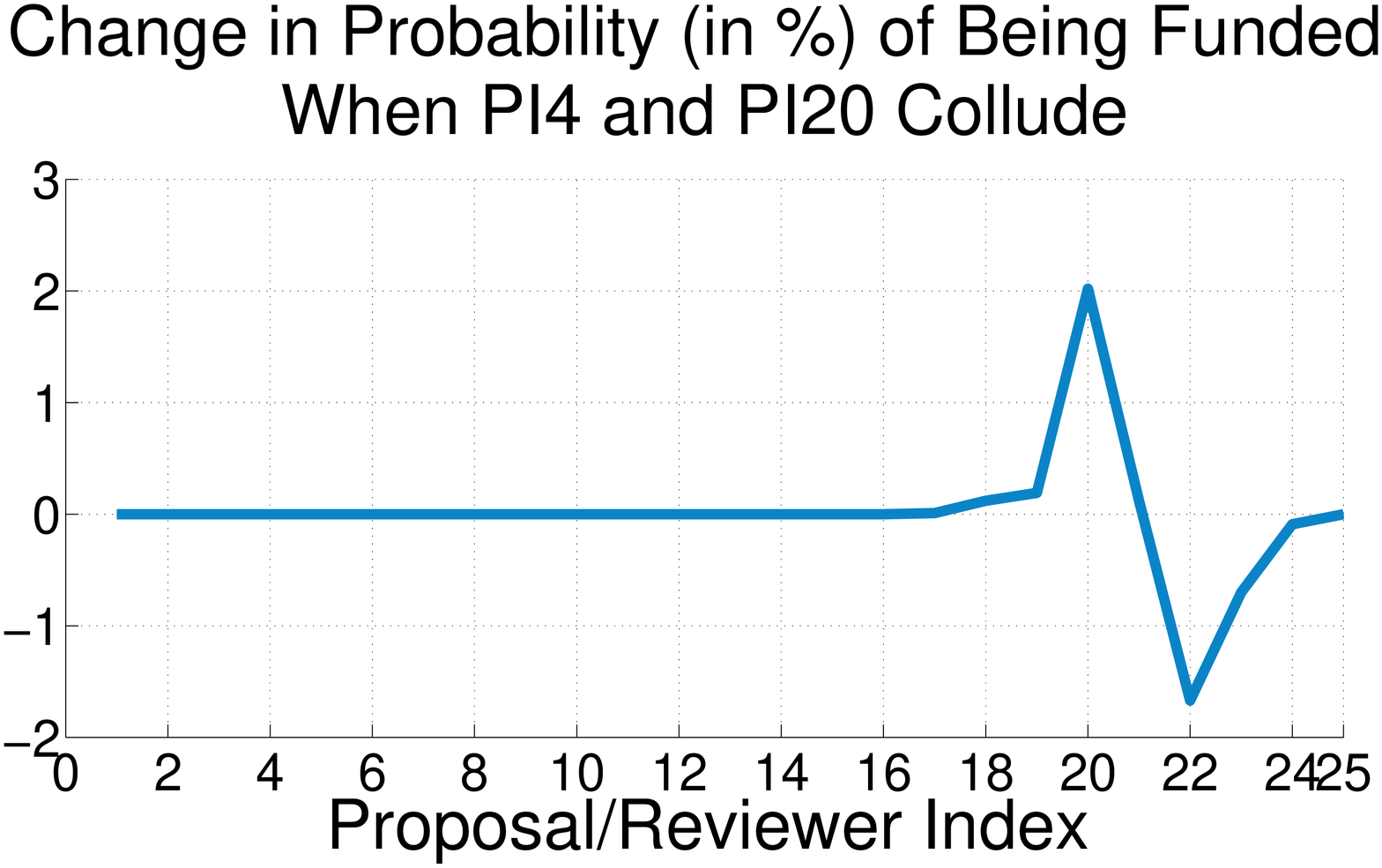}%
\caption{Reciprocal Favors - low quality and high quality}%
\label{R-HL}%
\end{minipage}
\hspace{0.5cm}
\begin{minipage}[t]{0.45\linewidth}
\includegraphics[width=\columnwidth]{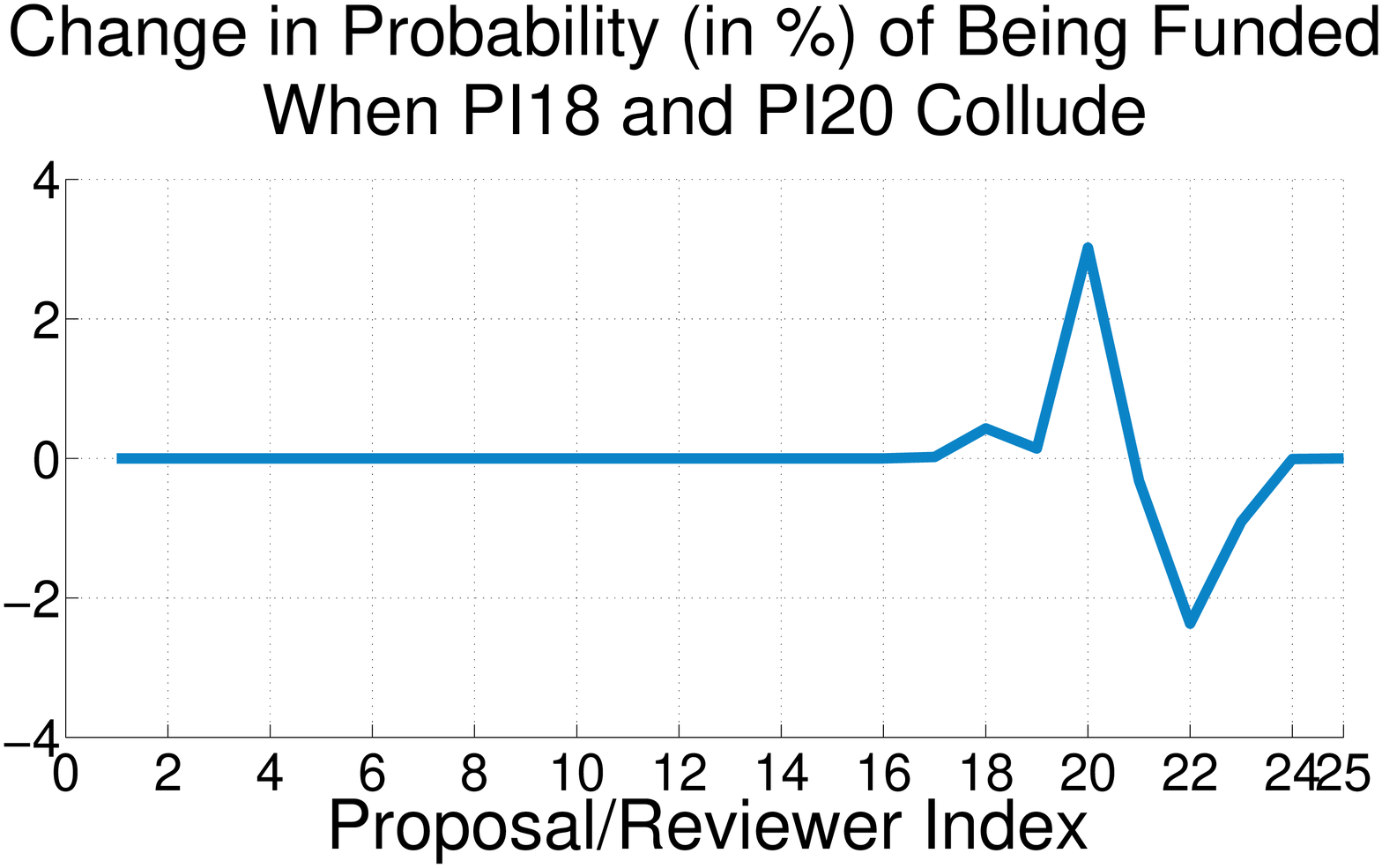}%
\caption{Reciprocal Favors - high quality proposals}%
\label{R-2H}%
\end{minipage}
\end{figure}

\textbf{Case I:} {One-Sided Favors - High/Low}.  
Assume PI\textsubscript{i} submits a low quality proposal, which has a low (or zero) chance of getting funded. Thus PI\textsubscript{i} has nothing to lose, be it with or without the quality bonus. In this case, the one-sided favors to PI\textsubscript{j} will increase {the latter's} chance of getting funded at the expense of lowering the chance of higher quality proposals, as shown in Fig. \ref{1S-HL}. An agreement between a high and a low quality proposals' PIs is a plausible scenario, in that applicants could actually arrange to have a low quality proposal submitted for the sole purpose of having a chance to promote a main proposal. Furthermore, the PI for a low quality proposal who has little prospect of being funded may be more open to such arrangement in return for future favors. 
%

\textbf{Case II:} {One-Sided Favors - High/High}. 
Assume PI\textsubscript{i} submits a proposal that is likely to get funded. However, by promoting PI\textsubscript{j}'s proposal, PI\textsubscript{i} is putting him/herself at risk of getting a lower score due to loss in quality bonus. As shown in Fig.  \ref{1S-2H}, this loss may be negligible. In fact, PI\textsubscript{18} decreases his/her chances of getting funded by roughly 0.3\% while increasing PI\textsubscript{20}'s chances by close to 30\%. 

\textbf{Case III:} {Reciprocal Favors - High/Low}.  
PI\textsubscript{j}, the PI for a high quality proposal, can further increase his/her chance of getting funded with reciprocal favors, but the gain is considerably lower than from Case I, since he/she is losing quality bonuses by promoting a low quality proposal, shown in Fig. \ref{R-HL}.  

\textbf{Case IV:} {Reciprocal Favors - High/High}. 
Finally, PIs with already relatively high quality proposals can collude to gain as shown in Fig. \ref{R-2H}. This means that the loss in quality bonus is not severe enough compared to the advantage gained from the collusion.\footnote{If even higher quality proposals collude, say PI\textsubscript{22} and PI\textsubscript{23}, they would both benefit but less than PI\textsubscript{20} did when colluding.  
Numerically, the probability of getting funded for PI\textsubscript{22} and PI\textsubscript{23} increases about 1\% and 0.5\% respectively, PI\textsubscript{24} and PI\textsubscript{20} are each down about 0.2\%, and PI\textsubscript{21} is down about 1\%.} 

Despite the benefits of quality bonuses discussed in Section \ref{sec:evil}, results here motivate the need for either an alternative incentive mechanism, or additional control over the reviewer assignment process, as the loss in quality bonus alone is inadequate to deter PIs from forming even the simplest coalitions. 


\subsection{Analysis}

We close this section by presenting a brief analysis based on the proposed utility model \eqref{eq:utility}, to illustrate the gains from colluding in one-sided favors from PI\textsubscript{i} to PI\textsubscript{j}. 

When all PIs report truthfully, $j$'s utility is derived in the same way as \eqref{eq:utility}. 
With one-sided favors, when $i\in {\cal R}_j$, the new expected MBC of proposal $j$ is given by: 
\[E[\text{MBC}_j] = \frac{j-1}{N-1} + \frac{N-j}{m(N-1)}~.\] 
The remaining proposals, if matched with reviewer $i$, expect higher MBCs if they are of lower quality, but are downgraded if they are of higher quality. 
Accordingly, the expected MBC of a proposal with index $k\neq i,j$ is given by:
\[E[\text{MBC}_k] = \frac{k-1}{N-1} - \frac{1}{(N-1)^2}(k-1-\frac{m-2}{m-1}(N-k))~.\] 
Finally, the quality bonuses that proposals $k\neq i$, including proposal $j$, receive are similarly affected by PI\textsubscript{i}'s altered report. Therefore, we will assume $E[B_j - B_k]=0,\ \forall k\neq i,j$. 

Using the above expressions, the expected utility of PI\textsubscript{j} when aided by PI\textsubscript{i} is given by: 
\begin{eqnarray}
E[u_j^{C}] 
 &=& E[u_j^{T}] + \underbrace{\frac{N-j}{m(N-1)} \sum_{k\neq j} \frac{1}{|j-k|^p}}_\text{Benefit from increased MBC}\notag\\
 && + \underbrace{\frac{1}{(N-1)^2}\sum_{k\neq j,i} \frac{k-1-\tfrac{m-2}{m-1}(N-k)}{|j-k|^p}}_\text{Benefit from decreased MBC of rivals} ~.
\label{eq:collude}
\end{eqnarray} 
As shown in \eqref{eq:collude}, the benefit from colluding with PI\textsubscript{i} is two-fold: it improves one's standing while possibly decreasing the chances of higher quality rivals. 

As observed earlier, PI\textsubscript{i} himself/herself may not be immediately benefiting from one-sided favors. This is mainly because PI\textsubscript{i} is losing quality bonuses. In addition, this PI may be further set back by reverse reporting, as he/she may end up promoting his/her low quality rivals. These effects are also observable using the utility function \eqref{eq:utility}: 
\begin{eqnarray}
E[u_i^{C}] 
 &=& E[u_i^{T}] + \underbrace{\frac{1}{(N-1)^2}\sum_{k\neq j,i} \frac{k-1-\tfrac{m-2}{m-1}(N-k)}{|i-k|^p}}_\text{Effect of reverse reporting}  \notag\\
 && - \underbrace{\frac{N-j}{m(N-1)|i-j|^p}}_\text{Loss due to j's increased MBC} - \underbrace{\sum_{k=1, k\neq i}^{N} \tfrac{E[B_i - B_k]}{|i-k|^p}}_\text{Loss in quality bonus}.
\label{eq:collude_2}
\end{eqnarray} 

%% file: Controversial.tex
\section{Controversial Proposals} \label{controversial}

Considering the significance of consistency between PIs' reviews in the proposed distributed review process, we devote this section to the subject of how the nature of the inaccuracy of reviewers and the disagreement among them on the more controversial proposals can affect the outcome of the review process, even when reviewers are somehow instructed to express their honest, yet differing viewpoints. 

\begin{figure}%
\centering
\includegraphics[width=0.8\columnwidth,height=0.15\textwidth]{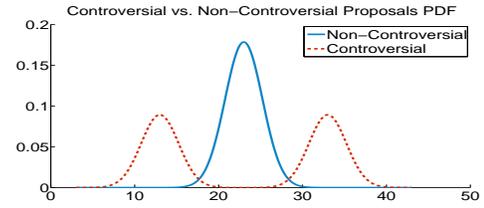}%
\caption{Controversial vs. non-controversial proposals} 
\label{bimodal}%
\end{figure}

To this end, we consider two classes of proposals, see Fig. \ref{bimodal}. \emph{Non-controversial} proposals are those for which reviewers' evaluations follow a Normal distribution, reflecting some noise in reviewers' assessments. In contrast, \emph{controversial} proposals follow a bi-modal distribution, modeling a possible disagreement on the intrinsic merit of the proposal. 

Specifically, consider a group of $N$ proposals indexed by their intrinsic merit. In general, each reviewer is able to evaluate a proposal subject to some error, such that reviewer $i$ evaluates proposal $j$ as $r_{ij} = j + b_{ij}$. Reviewer $i$ then sorts the proposals in ${\cal A}_i$ based on $r_{ij}$ and assigns Borda scores of $0$ to $m-1$ accordingly. 
We assume only one of the proposals, $k$, is controversial, so that for each $i\in {\cal R}_k$, $b_{ik}\sim {\cal N}(\mu_{ik},\sigma^2)$, $\mu_{ik}$ is either $+\delta\mu$ or $-\delta\mu$ with equal probability. The remaining proposals $j\neq k$ are non-controversial, with $b_{ij}\sim {\cal N}(0,\sigma^2)$. 

\begin{figure}%
\begin{minipage}[t]{0.45\linewidth}
\includegraphics[width=\columnwidth]{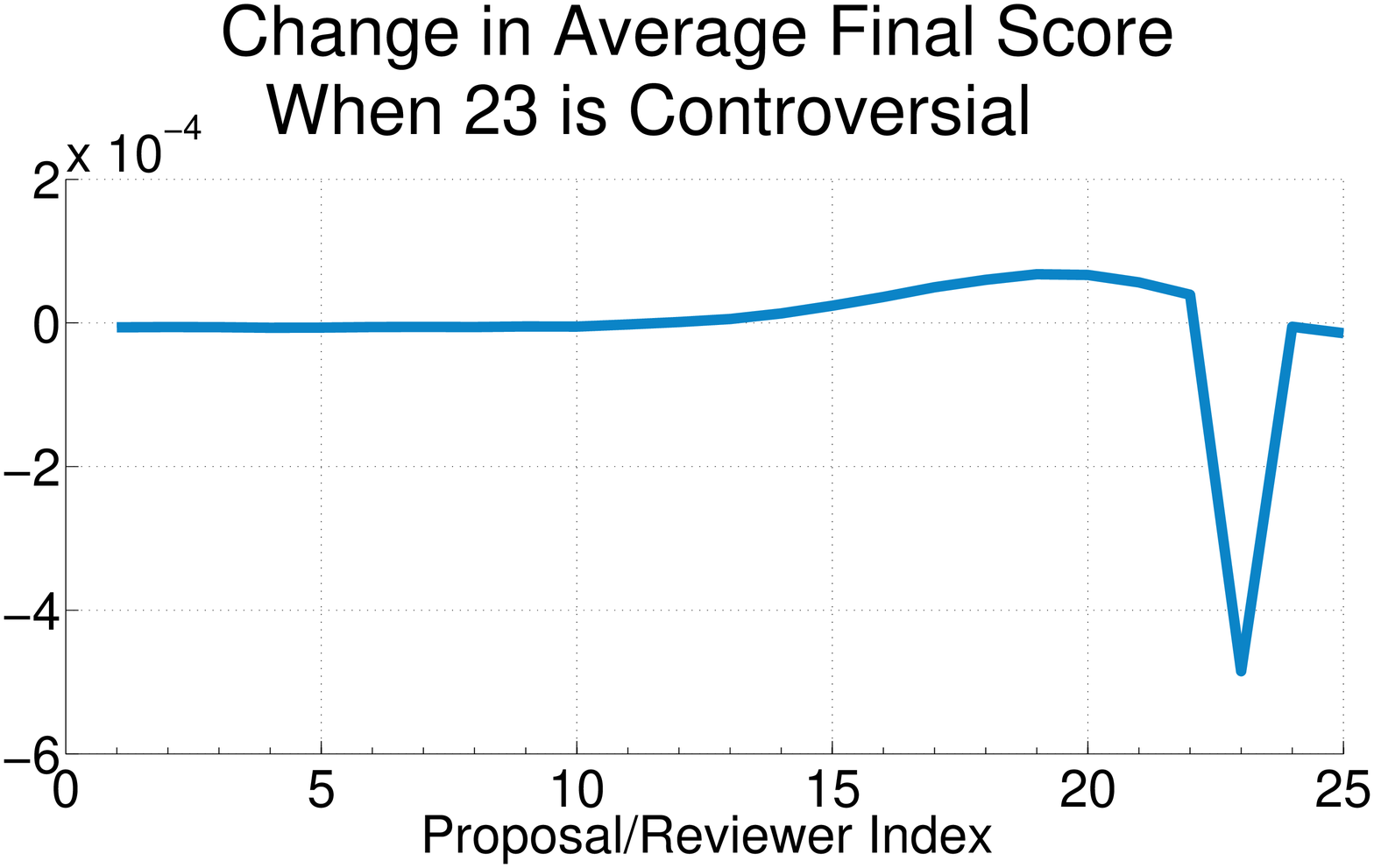}%
\caption{A controversial proposal looks worse than its true/intrinsic value}%
\label{accuracy4}%
\end{minipage}
\hspace{0.4cm}
\begin{minipage}[t]{0.45\linewidth}
\includegraphics[width=\columnwidth]{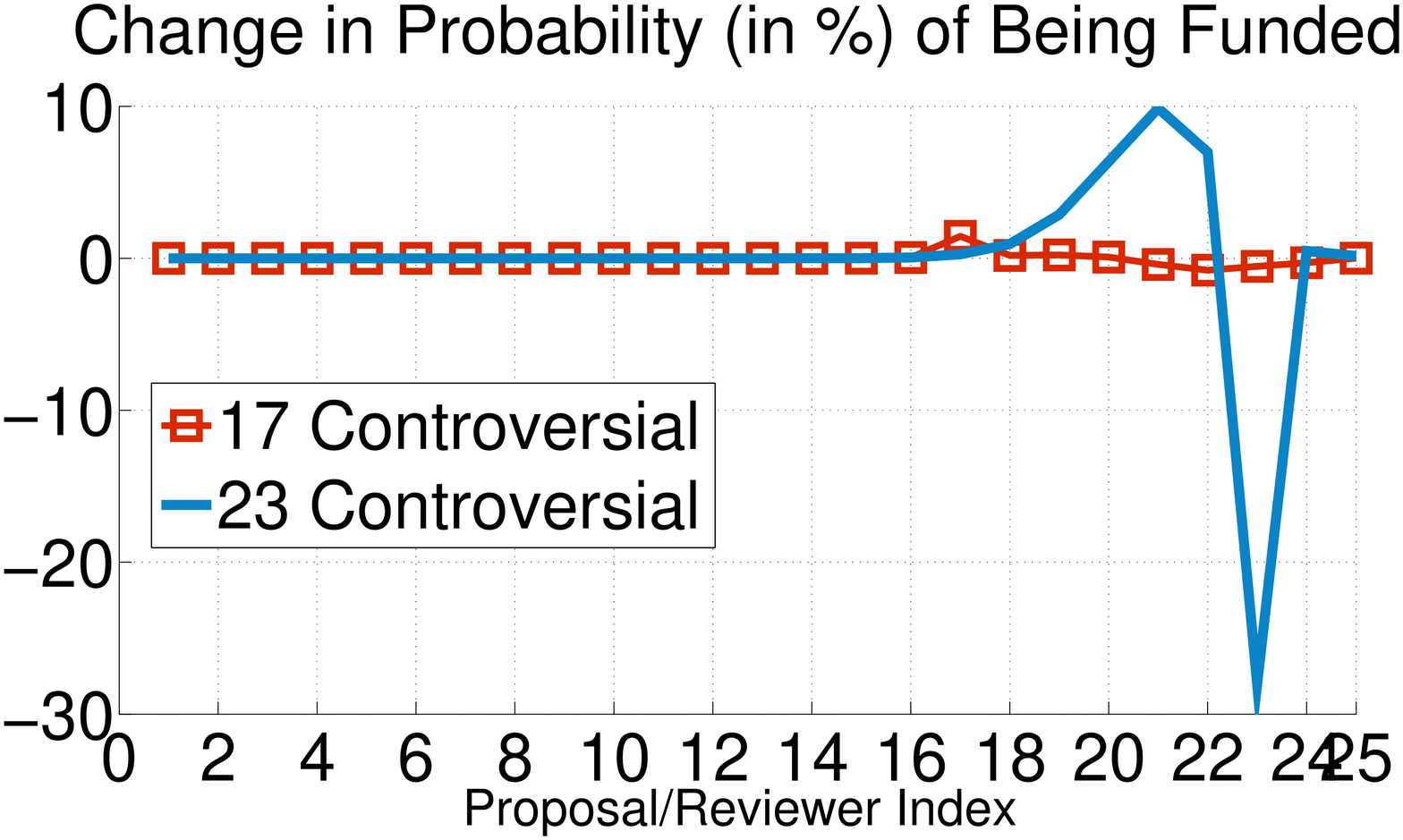}%
\caption{Controversial w/ low quality benefit, high quality lose}%
\label{accuracy3}%
\end{minipage}
\end{figure}

For the following simulations, we set $\sigma = 2.5$ and $\delta\mu = 5$. A choice of $\sigma = 2.5$ means that reviewers are able to evaluate a proposal within 5 places of its intrinsic merit with high probability.  
The first natural expectation is that controversial proposals are likely to be at a disadvantage compared to their non-controversial close rivals due to their set of mixed reviews. This is verified in Fig. \ref{accuracy4}, where a controversial proposal of intrinsic merit 23 looks worse than its true value. {\em Nevertheless, our results show that this speculation is only half true -- controversy appears to actually help intrinsically lower quality proposal, but hurt intrinsically high-quality proposals,} as shown in Fig. \ref{accuracy3}. To understand this, consider a proposal of merit 17, which would not be normally funded. By receiving positive reviews from some reviewers, this proposal will overall get a leg up in the ranking.  A high quality proposal with true merit 23 on the other hand has a lower chance of getting funded when it is controversial. This is because higher ratings are not likely to change the MBC of this proposal, as such proposal should already stand out against other competitors in the pile. Low ratings (due to the controversy) on the other hand can bring down the MBC of this proposal, putting it at an overall disadvantage. {Consequently, this review process will not only tend to disfavor controversial proposals, much as expected, but also tend to favor mediocrity among equally controversial proposals}. 

Finally, note that the reviewers of controversial proposals are also at a disadvantage. These reviewers will receive lower quality bonuses as a result of disagreeing with one another, and are set back by the reviewers lucky enough not to have been handed any controversial proposals, as the latter are more likely to be in sync with the global view, and thus receive high quality bonuses.

%% file: Discussion.tex
\section{Discussion and Conclusion} \label{disc}

We have focused on three specific aspects of the new NSF SSS program review pilot procedure, and highlighted some of its strengths and limitations. 

\begin{figure}%
\centering
\includegraphics[width=0.8\columnwidth]{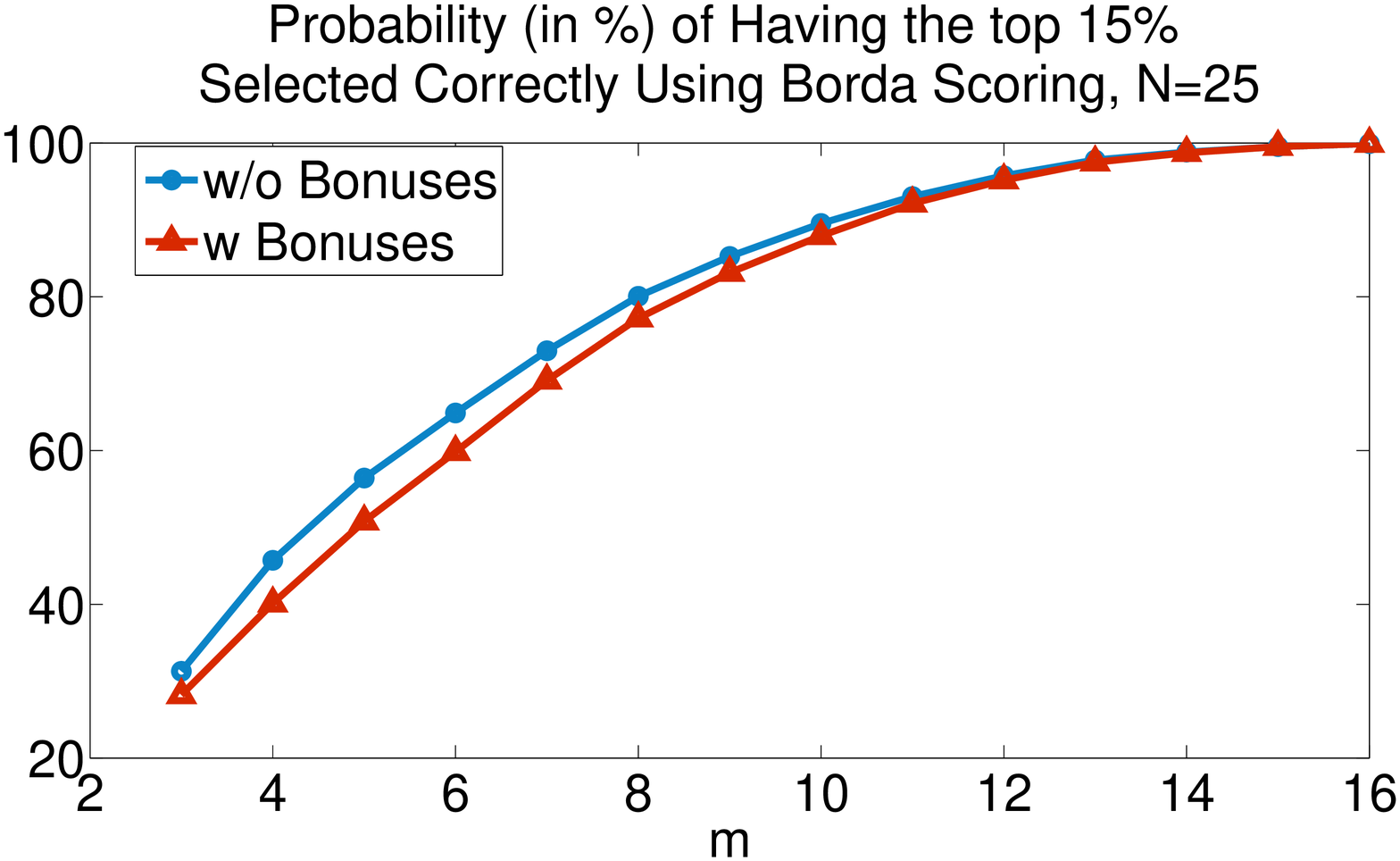}%
\caption{Effect of Increasing $m$}%
\label{inc_m}%
\end{figure}

On the first concern of whether the quality of reviews should be part of the proposals' final assessment, as its benefits in preventing dishonest behavior are more considerable than the possible losses, we conclude that the use of quality bonuses as a commodity of interest is a viable option. 
Though discrimination against less competent reviewers is inevitable, some of the inaccuracies induced by the use of quality bonus may be alleviated as follows. As pointed out in Section \ref{Bonus}, despite being honest, accurate, and competent, some PIs may receive lower bonuses depending on the outcome of the review assignment process. This effect can be avoided by assigning a more or less uniformly distributed set of proposals to each reviewer, either using a pre-processing round, two rounds of reviews, or by increasing $m$, the number of proposals assigned to each reviewer \cite{blog3}. This latter solution will increase the accuracy of the review process, while reducing this side effect of including quality bonuses, see Fig. \ref{inc_m}, though obviously it also increases the burden on the participants.

Alternatively, both drawbacks could be avoided by using the quality of reviews as a numeraire commodity. For instance, imposing limits on future submissions and decreasing the workload of future review assignments, are two extreme examples of negative and positive numeraire bonuses, respectively. However, as the valuation of these commodities is often difficult to assess, their implementation will be more challenging. 

The issue of collusion may be ameliorated by collecting, {\em a priori}, CoI lists, i.e., before exposing the list of PIs in a group, {or, by limiting the review assignments to those such that no two PIs are directly matched (reviewing each other's proposal)}. Also, a final round of reviews of the higher ranked proposals by a panel of experts could help weed out proposals that have been unjustly promoted. 

Dealing with controversial proposals may be more challenging, as the inclination of the review system towards non-controversial proposals is the result of multiple factors, including the nature of Borda scoring, and the applicants' and reviewers' disinterest in high-risk proposals. A possible solution would be to explicitly encourage innovative proposals through the initial call for proposals and review guidelines \cite{Merrifield09, blog2}, as is done currently. In addition, it is possible to detect controversial proposals based on their diverse set of reviews. As the number of these proposals is often small, it might be reasonable to suggest further discussion by the set of reviewers, or adding a separate round of reviews, either by a different group of PIs, or by a panel of expert reviewers.  Also, adding a comment phase before voting, in which reviewers can share (anonymized) comments with the other reviewers, may bring attention to innovative proposals \cite{blog3}. 
Finally, by enlarging the {message space}, reviewers can be given an additional entry to specifically present their views about high risk proposals  \cite{blog2}. 


Lastly, we note that in addition to scientific journals and funding committees, the distributed mechanism can be applicable in other peer-review settings, e.g. medical peer-reviews, where the built-in incentives could lessen concerns of {sham peer-reviews} \cite{sham}. 
 Furthermore, a similar approach can be implemented in elicitation and peer-prediction problems, which are used to aggregate users' predictions about a given event \cite{goel09}. A main assumption in many of the existing mechanisms is that users do not attach any value to the outcome the elicitor is building using the aggregated data. The proposed method could prove especially useful if this assumption fails to hold, e.g. when aggregating reputation-related data from a group of competitors.